\documentclass[preprint,brazil,english,sort&compress,authoryear]{elsarticle}
\usepackage[T1]{fontenc} \usepackage[utf8]{inputenc} \usepackage[brazil]{babel}
\usepackage[unicode=true,
bookmarks=true,bookmarksnumbered=false,bookmarksopen=false,
breaklinks=false,pdfborder={0 0 1},backref=false,colorlinks=false,hidelinks]
{hyperref} \hypersetup{pdftitle={Automated pulse discrimination of two
freely-swimming weakly electric fish and analysis of their electrical behavior
during a dominance contest}, pdfauthor={Rafael Tuma Guariento, Thiago Schiavo
Mosqueiro, Paulo Matias, Vinicius B Cesarino, Lirio Onofre Baptista de Almeida,
Jan Frans Willem Slaets, Leonardo Paulo Maia, Reynaldo Daniel Pinto},
pdfsubject={Signal processing and machine learning supporting neuroethology
research}, pdfkeywords={Neuroethology,Electrocommunication,Inter Pulse
Interval,Pulse Discrimination,Chirp Detection,Off,Random Forest,Dominance
Contest,Submissive,Gymnotus,Pulse Type Electric fish}}

\usepackage{mathtools}

\usepackage{graphicx} \graphicspath{ {/}{figs/} }

\usepackage{hyperref}

 \makeatletter

 \journal{Journal Physiology Paris}

 \biboptions{compress}

 \makeatother

\begin{document}

\title{Automated pulse discrimination of two freely-swimming weakly electric
fish and analysis of their electrical behavior during a dominance contest}

\author[ifsc]{Rafael~T~Guariento\corref{corauthor}}
\ead{rafael.tuma.guariento@usp.br}

\author[ucsd]{Thiago~S~Mosqueiro}

\author[ifsc]{Paulo~Matias}

\author[ifsc]{Vinicius~B~Cesarino}

\author[ifsc]{Lirio~OB~Almeida}

\author[ifsc]{Jan~FW~Slaets}

\author[ifsc]{Leonardo~P~Maia}

\author[ifsc]{Reynaldo~D~Pinto}

\cortext[corauthor]{Corresponding author. Tel.: +55 16 33738090; Fax: +55 16
33739879}

\address[ifsc]{\foreignlanguage{brazil}{São Carlos} Institute of Physics,
University of \foreignlanguage{brazil}{São Paulo}, PO Box 369, 13560-970, São
Carlos, SP, Brazil.}

\address[ucsd]{BioCircuits Institute, University of California San Diego, CA,
USA}

\date{June 2016}

\begin{abstract}


Electric fishes modulate their electric organ discharges with a remarkable
variability. Some patterns can be easily identified, such as pulse rate
changes, offs and chirps, which are often associated with important behavioral
contexts, including aggression, hiding and mating. However, these behaviors are
only observed when at least two fish are freely interacting. Although their
electrical pulses can be easily recorded by non-invasive techniques,
discriminating the emitter of each pulse is challenging when physically similar
fish are allowed to freely move and interact. Here we optimized a custom-made
software recently designed to identify the emitter of pulses by using automated
chirp detection, adaptive threshold for pulse detection and slightly changing
how the recorded signals are integrated. With these optimizations, 
we performed a quantitative analysis of the statistical changes throughout the 
dominance contest with respect to Inter Pulse Intervals, Chirps and Offs
dyads of freely moving \textit{Gymnotus carapo}. In all dyads, chirps 
were signatures of subsequent
submission, even when they occurred early in the contest.  Although offs were
observed in both dominant and submissive fish, they were substantially more
frequent in submissive individuals, in agreement with the idea from previous
studies that offs are electric cues of submission. In general, after the
dominance is established the submissive fish significantly changes its average
pulse rate, while the pulse rate of the dominant remained unchanged.
Additionally, no chirps or offs were observed when two fish were manually kept
in direct physical contact, suggesting that these electric behaviors are not
automatic responses to physical contact.

\end{abstract}


\maketitle

\section{Introduction}

Social hierarchies are established and maintained by a broad range of dynamic
behaviors expressed by animals during communication \citep{lorenz}.  Weakly
electric fish in the genus \textit{Gymnotus} are territorial and show marked changes in
their motor and electrical behavior once a dominance hierarchy is established
\citep{westbyComparative, westbyFuther, anaSilvaNonSex}. The dominant fish
actively swims and explores the whole environment, while the submissive fish
often remains still. Sequences of Inter Pulse Intervals (IPI) reveal highly
variable patterns that are clearly distinct before and after the dominance
contest \citep{westbyComparative, anaSilvaNonSex, anaSilva2015}. Two critical
tasks are performed by using self- and conspecifc-generated electrical pulses
and their feedback on fish's electrosensory system: electrolocation and
electrocommunication \citep{vonderEmde2013, Caputi2008, black1970role,
castelloJeb}. Detecting distortions on the stereotyped self-generated electric
pulse provides information from the environment\citep{jun2014, Pereira2010},
while IPI patterns are likely used for communication among conspecifics
\citep{forlimPinto}.

While fish are contesting dominance, they may stop emitting pulses during
variable time intervals ("offs") \citep{westbyComparative, westbyFuther}.  In
some situations, instead of the stereotyped pulses, these fish can emit small
electric field oscillations, known as "chirps” (Figure \ref{fig:chirp})
\citep{anaSilva2015, anaSilvaNonSex}.  Offs and chirps are often related to
submission \citep{anaSilvaNonSex}, physical aggression and retreat. Therefore,
they might be important flags used by fish to convey submission or stress. There
is strong evidence that electrocommunication has an important role in
dominance definition and maintenance \citep{westbyComparative, westbyMcGregor}.

However, to assess electrocommunication, one of the main challenges is to
discriminate pulses from freely interacting fish \citep{black1970role,
westbyMcGregor, Letelier}. Two distinct but electrically coupled aquaria have
been used to avoid this problem \citep{forlimPinto}, but complex behaviors such
as chirps have never been detected with such artificial setups, possibly due
the lack of behavioral cues other than the electric pulses (e.g.,movement and
bites).

Here we report on improving a state-of-art classification technique
\citep{Matias}, that discriminates pulses from \textit{Gymnotus carapo} dyads,
to allow automated chirp detection. We used a supervised learning algorithm,
and its training required only short samples of time series with and without
chirps.  We applied these tools to analyze the electrical interactions during
and after dominance interaction. Fish dominance roles were identified  by
observing behavioral cues. We analyzed data from several dyads, and discuss the
changes found in the distributions of IPIs, offs and chirps, during their
dominance contest. Because chirps often occur when both fish engage in
close-range physical contact, we tested the hypothesis that chirps and offs
could be automatically generated by the contact of electric organs when fish
are touching each other. However, no chirps nor offs were observed when two
fish were manually placed in physical contact, with their skins touching.  This
suggests that instead of related to an automatic mechanism, chirps and offs
might be used to communicate important information that shapes the dominance
context.

All our software is freely available \citep{gymnotools}. Data from one of our dyads is also available \citep{dados}.

\section{Materials and Methods}

\subsection{Ethics statement}

All experimental protocols and procedures were in accordance with the ethical
principles of the Society for Neuroscience and were approved by the Committee
on Ethics in Animal Experimentation of the São Carlos Institute of Physics --
University of São Paulo.

\subsection{Subjects and housing}

Experiments were conducted on $6$ healthy adult specimens of \textit{Gymnotus
carapo}, $15$--$25$~cm long, regardless of sex. The home tanks and feeding
protocol were previously described \citep{forlimPinto}. All specimens were
acquired from local commercial suppliers within 15 days before experiments.
Four specimens were acquired on September/October 2015 and two on June 2016.

\subsection{Experimental setup}

The experiments with freely interacting fish were performed in a glass aquarium
\( (100\times50\times50\,\mathrm{cm} ) \) filled with tap water, and shielded
by a grounded metallic mesh (Faraday cage). To induce interaction between pairs
of fish, no objects were placed in the measurement aquarium, leaving no spots
for hiding. The water was set at room temperature \((23 \pm 2)
\mathrm{^{\circ}C}\) and the conductivity was measured before and after the
experiments as \((55 \pm  5) \mathrm{\mu S / cm}\). The fish were placed in
this setup only during the experiments.

\begin{figure} \centering \includegraphics[width=1.\textwidth]{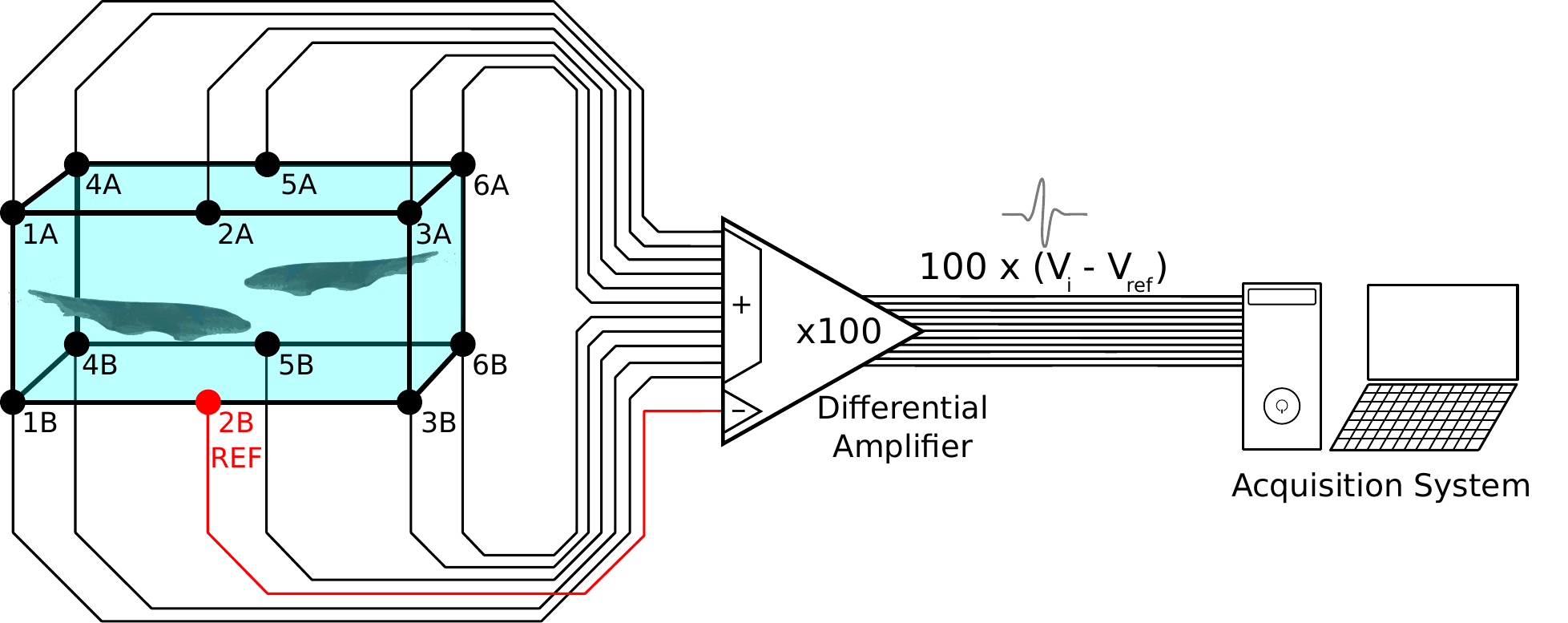}
\caption{The fish are allowed to move freely and interact in the measurement
aquarium, where the discharges from their electric organ were recorded. The
time series from $11$ electrodes were differentially amplified $100$ times,
using one electrode as reference (red wire in the figure), and then acquired by
a commercial analog-to-digital converter at \(45.5~kHz\).} \label{fig:inst_aq}
\end{figure}

The Electric Organ Discharges (EODs) were measured using a three-dimensional
array of $12$ electrodes, each consisting of a \(0.2 \mathrm{mm}\) diameter
stainless steel wire (Figure \ref{fig:inst_aq}). The electrodes were inserted
through the vertices and in the mid edge of the longer sides of the measurement
aquarium.  Time series of 11 electrodes were differentially amplified (100
times
- Texas Instruments Operational Amplifier TL07X series on inverting mode with a
  10Hz input high-pass filter) with a single common reference electrode, and
digitized at \(45.5~\mathrm{kHz}\) by a commercial acquisition system (Digidata
1322A, Molecular Devices).

\subsection{Time series from isolated fish and training protocol}
\label{sec:training}

To collect enough data to calculate pulse shape statistics over a wide range of
positions, each fish was placed alone in the measurement aquarium to record
their own EOD time series. Each fish was left freely swimming for 10 minutes
and then induced to swim for another 10 minutes, by prodding the fish with a
non-conductive net.  These time series were then used as training examples for
a protocol based on state-of-art machine learning techniques \citep{Matias}, as
described in section \ref{sec:discrimination}.

\subsection{Dominance contest}

Each pair of fish was placed simultaneously in the measurement aquarium to
interact for $70$min, when a contest for dominance happened, and then, moved
back to their home tanks. All the experiments were performed at night (2000 -
0200h) in the dark. According to classical signatures described in the literature
\citep{westbyComparative, anaSilvaNonSex, anaSilva2015}, after each experiment a 
dominance relationship was formed. For instance, one of the fish in each dyad 
rapidly stop biting its conspecific and frequently swam away from their aggressive 
counterparts. These fish were later identified as submissive. 
The dominance contests were performed \textit{before} the training
protocol as defined by section \ref{sec:training}, assuring that the animals
were naïve to the experimental setup.

\subsection{Chirp detection} \label{sec:chirp}

\begin{figure} \centering \includegraphics[width=1.2\textwidth]{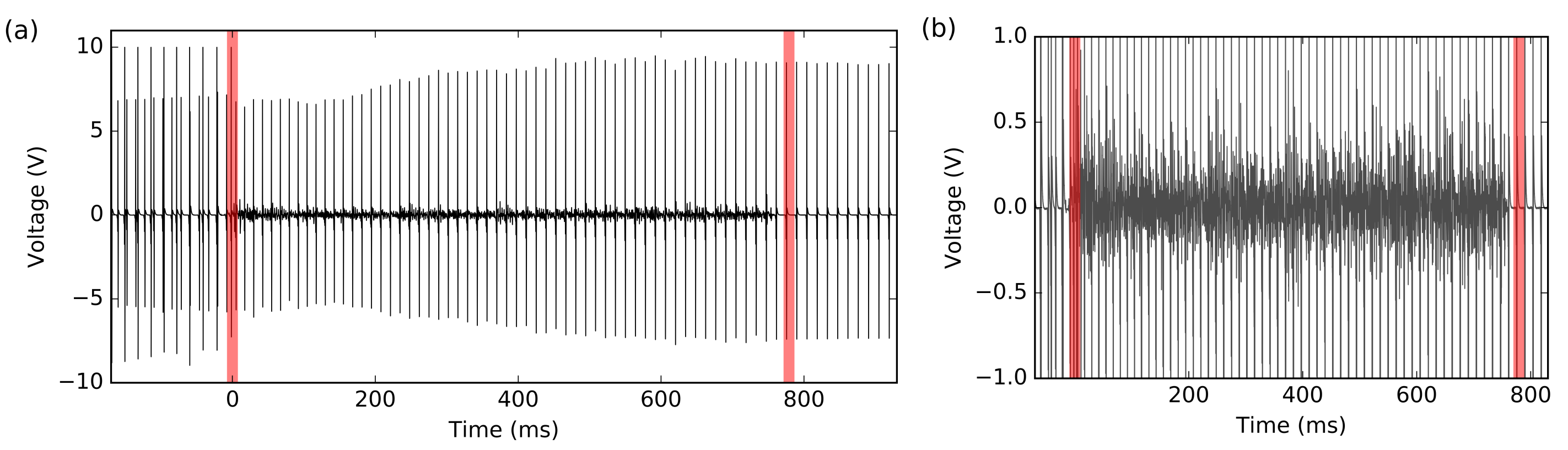}
\caption{Time series from a single electrode containing a chirp.  {\bf (a)} One
of the fish stops firing at $t=0$ and emits noisy-like oscillations with zero
mean and amplitude ten times smaller than normal pulses (compare with $t<0$).
{\bf (b)} Zoom on chirp signal.} \label{fig:chirp} \end{figure}

We used supervised learning to detect chirps in the time series
(Figure \ref{fig:chirp}a,b). First, we built a new time series by summing the
absolute values of voltage of all electrodes. Several non-overlaping sections
were uniformly selected, each one $0.5$~s long, and manually inspected and
labeled according to two classes: (i) sections with chirp and (ii) sections
without chirps. Then, $N$ sections with chirp and $2N$ without chirp were fed
as training examples for a Random Forest \citep{randomForest} classifier. To
implement this classifier as a continuous chirp detector, the training examples
were segmented in sampling windows of $2000$ data points ($44$~ms). The
accuracy of the classification was measured by $5$-fold cross validation
\citep{bishop2006pattern} and increased with $N$.

Empirically, chirps were always longer than $100$ms, i.e. $3$ moving windows of
$44$~ms.  To avoid misclassification of smaller regions, when applying the
detector on the whole timeseries, chirp detection was only considered when $4$
consecutive windows of $2000$ data points ($44$~ms) were classified as
containing chirps (Figure \ref{fig:chirp}c).  Similarly, the end of the chirp
was considered when $10$ consecutive windows were classified as not containing
chirps.

All chirp detection programs were written in Python and are freely available
GitHub \citep{gymnotools}.

\subsection{Fish discrimination} \label{sec:discrimination}

Each pulse was assigned to a fish by using supervised learning, based on
previous methodology \citep{Matias} with minor improvements. Specifically, we
applied Hilbert transform to the time series of all $11$ electrodes, summed
their absolute values, and identified peaks. In regions without chirps, the
position of each pulse was then detected by finding the peaks of the summed
signal with a threshold of $1 V$. In regions containing chirp, the threshold
was adaptively set as 30\% of the summed signals of the maximum value in the
last $300 ms$. The timing of each pulse is defined when the maximum at each
peak occurred.  Using this method, pulse detection is independent on the fish
position in the measurement aquarium \citep{jun2014}. We also implemented a
Graphical User Interface (GUI) to verify and manually correct the few pulses
wrongly classified. All the discrimination programs were written in C++ and
Python, and are freely available on GitHub \citep{gymnotools}.

\subsection{Statistical tests on time series of Inter Pulse Interval}

To test whether distributions of Inter Pulse Intervals (IPIs) changed
throughout the dominance contest, we applied a paired Student's $t-test$ on
a subsampled version of the time series to avoid pseudoreplication
\citep{lazic2010problem}. IPI time series were segmented using
sections of $1500$ points, which represents about $18$s on average.
Because this is larger than the average correlation length ($300 \pm 200$), each sample
is mostly uncorrelated from each other. Because the distribution of
IPIs in most cases resemble a bell-shaped distribution (see Figure S6)
and because failing the hypothesis of normally distributed samples
induces less errors than the existence of correlation among samples
\citep{lazic2010problem}, we used a paired $t-test$. For completeness,
we show all $t$ statistics and p-values on table S2 of Supplemental Material.

\subsection{Response of electric organs by physical contact}

To investigate whether offs and chirps are automatically generated by physical
contact, causing some sort of electrical interference or short-circuiting, we
picked pairs of fish and placed them by hands, until their skins
touched. We repeated this experiment $4$ times with distinct pairs, both using
bare hands and gloves, but no difference was observed.  Their bodies were
positioned for $2-5$ minutes in parallel, anti-parallel and orthogonal
directions (Figure \ref{fig:esfrega} Top), with an interval of about $2$
minutes between positions.  A dipole electrode was placed near each fish to
record its electric activity, using commercial apparatus (A-M Systems 1700
differential AC amplifier, Digidata 1322A, Molecular Devices).

\section{Results}

\begin{figure} \centering
\includegraphics[width=1.15\textwidth]{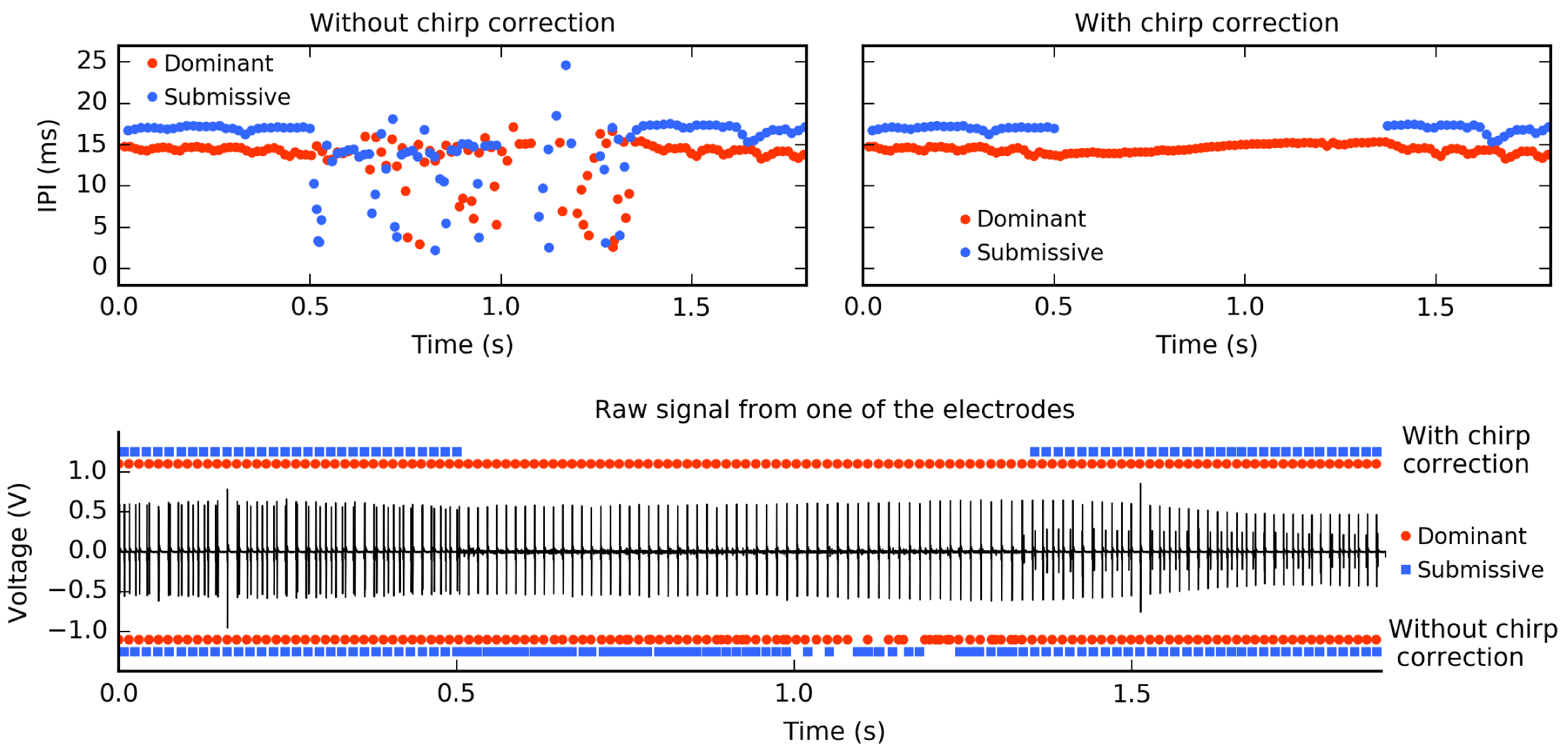}

  \caption{Fish discrimination was improved by automatically detecting chirp
regions and changing the threshold for spike detection on chirp regions.
\textbf{Top left:} Time series of Inter Pulse Intervals (IPI) before the chirp
detection. \textbf{Top right:} Time series of  Inter Pulse Intervals (IPI) of
the same region after the chirp detection. \textbf{Bottom:} Time series
acquired by one of the electrodes and pulse discrimination without (below) and
with (above) chirp detection and threshold adaptation.}

  \label{fig:chirp_importance}
\end{figure}



\begin{figure}[t] \centering \includegraphics[width=1.2\textwidth]{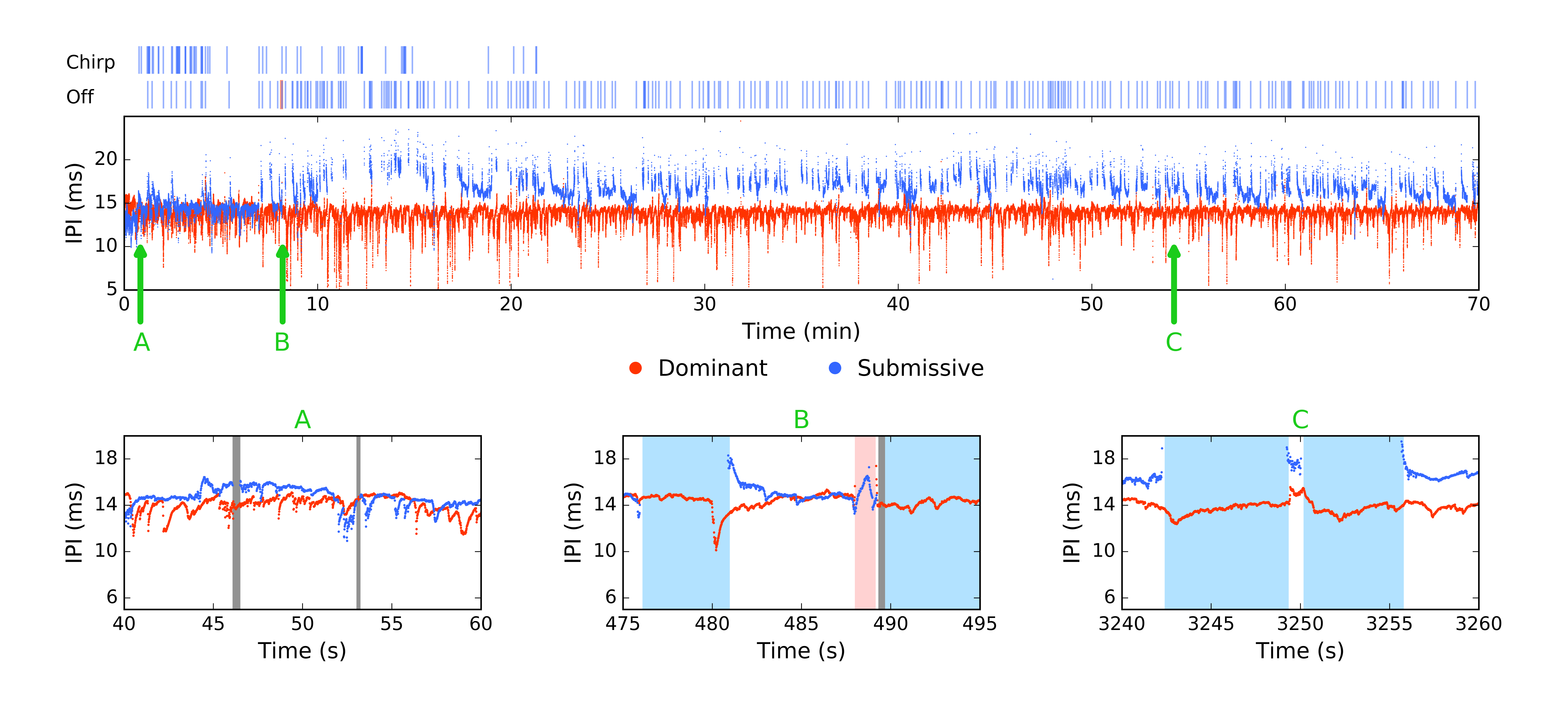}

  \caption{Time series of Inter Pulse Intervals (IPIs) with two fish freely
swimming in the same aquarium. Blue represents the submissive fish and red represents
the dominent fish. \textbf{Top:}~Complete recording with bars
representing the time stamps of the onset of chirps and offs. The gaps on the
IPI time series are due to chirps and offs. The timestamps of each off and chirp
are represented by vertical bars above the plot.
\textbf{Bottom:}~Zoom on the IPI time series. Chirps are represented as gray regions. 
Offs are represented as blue regions for the submissive fish and red regions for the dominant
fish. \textbf{Bottom A:}~During the early dominance 
contest ($t < 430s$).
\textbf{Bottom B:}~During the dominance contest ($430s < t < 1000s$).
\textbf{Bottom C:}~After the dominance contest ($t > 1000s$).}

  \label{fig:series}

\end{figure}

\begin{figure}[t] \centering
\includegraphics[width=0.90\textwidth]{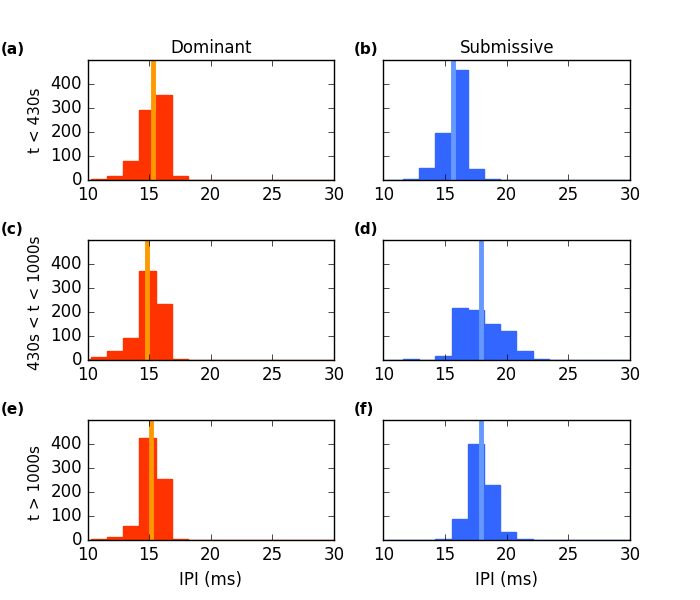} \caption{IPI
Histograms for IPIs smaller than $30$~ms (without chirps and offs) for dyad 1. Vertical
bars denote averages. \textbf{Left:} No significant changes were observed for
the dominant fish (Mann-Whitney U, $p > 5$\%). \textbf{Right:} Submissive fish
changes to a broader distribution and eventually stabilizes at a significantly
larger average (one-tail Mann-Whitney U, $p < 2.5$\%). \textbf{(a) and (b):}
Histograms for the first 430 seconds. \textbf{(c) and (d):} Histograms for time
between 430 and 1000 seconds. \textbf{(e) and (f):} Histograms for times
greater than 1000 seconds.} \label{fig:histFine}

\end{figure}

\begin{figure} \centering
\includegraphics[width=1.1\textwidth]{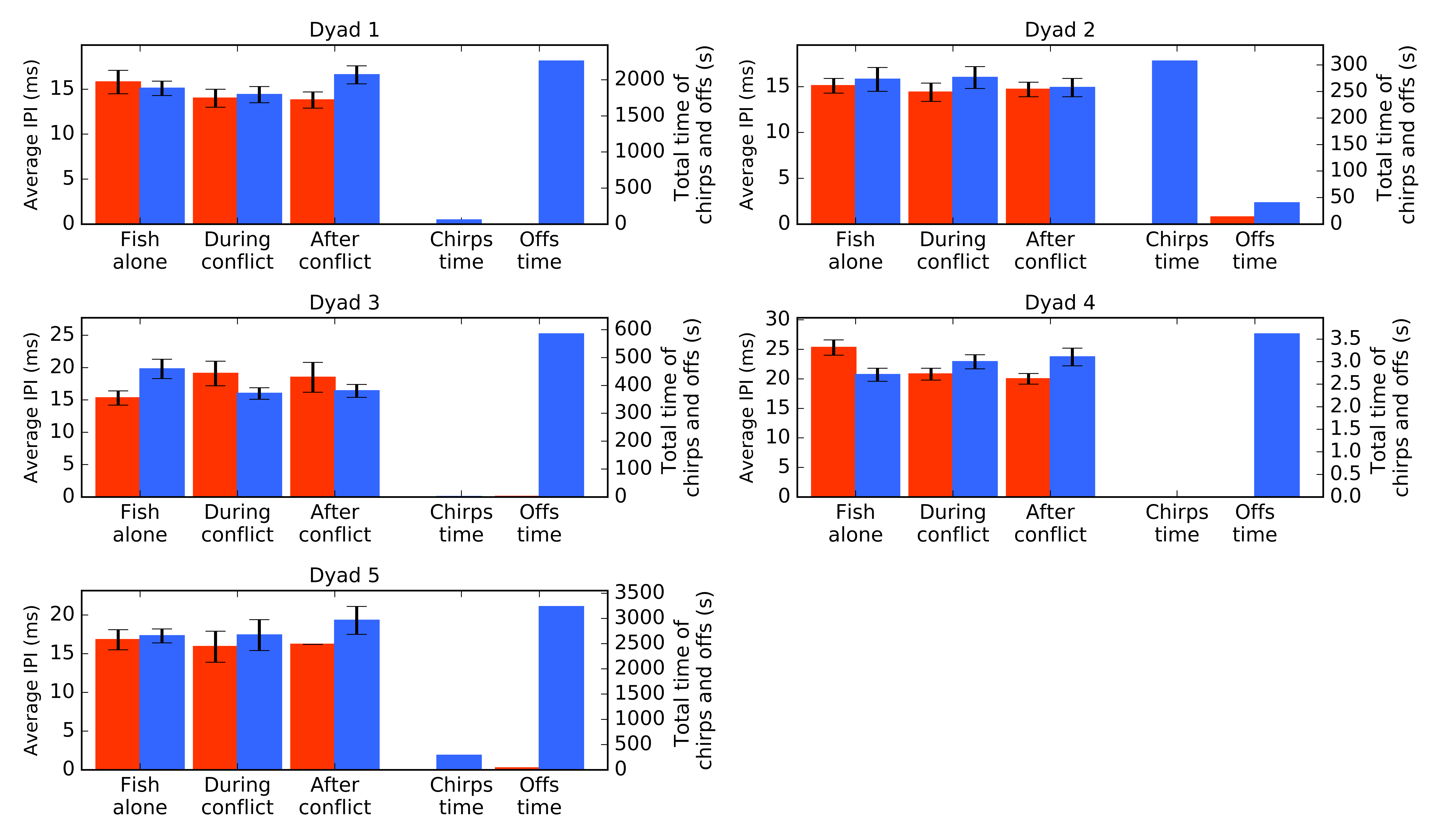}

  \caption{Average and standard deviation of Inter Pulse Intervals (IPI) and
occurrences of chirps and offs for all dyads, except dyad 6 (See Supplemental
Material Tables S1 and S2, and Figures S3 and S6). To not be influenced by offs and
chirps, the mean and standard deviation of IPIs were measured only of IPIs
shorter than $40$ms. Offs were defined as IPIs with more than $1$s. 
The sections "Fish alone", "During contest" and "After conflict" correspond, 
respectively, to the training protocol (See section \ref{sec:training}), to 
the first $430$s of the interaction 
between both fish, and the interaction after $1000$s.
Mean and standard deviation values can be found on Supplemental Material. A 
plot of the IPI series for each dyad can also be found on the Supplemental 
Material.}

  \label{fig:resultsallpairs} \end{figure}

\begin{figure}[t] \centering
\includegraphics[width=0.9\textwidth]{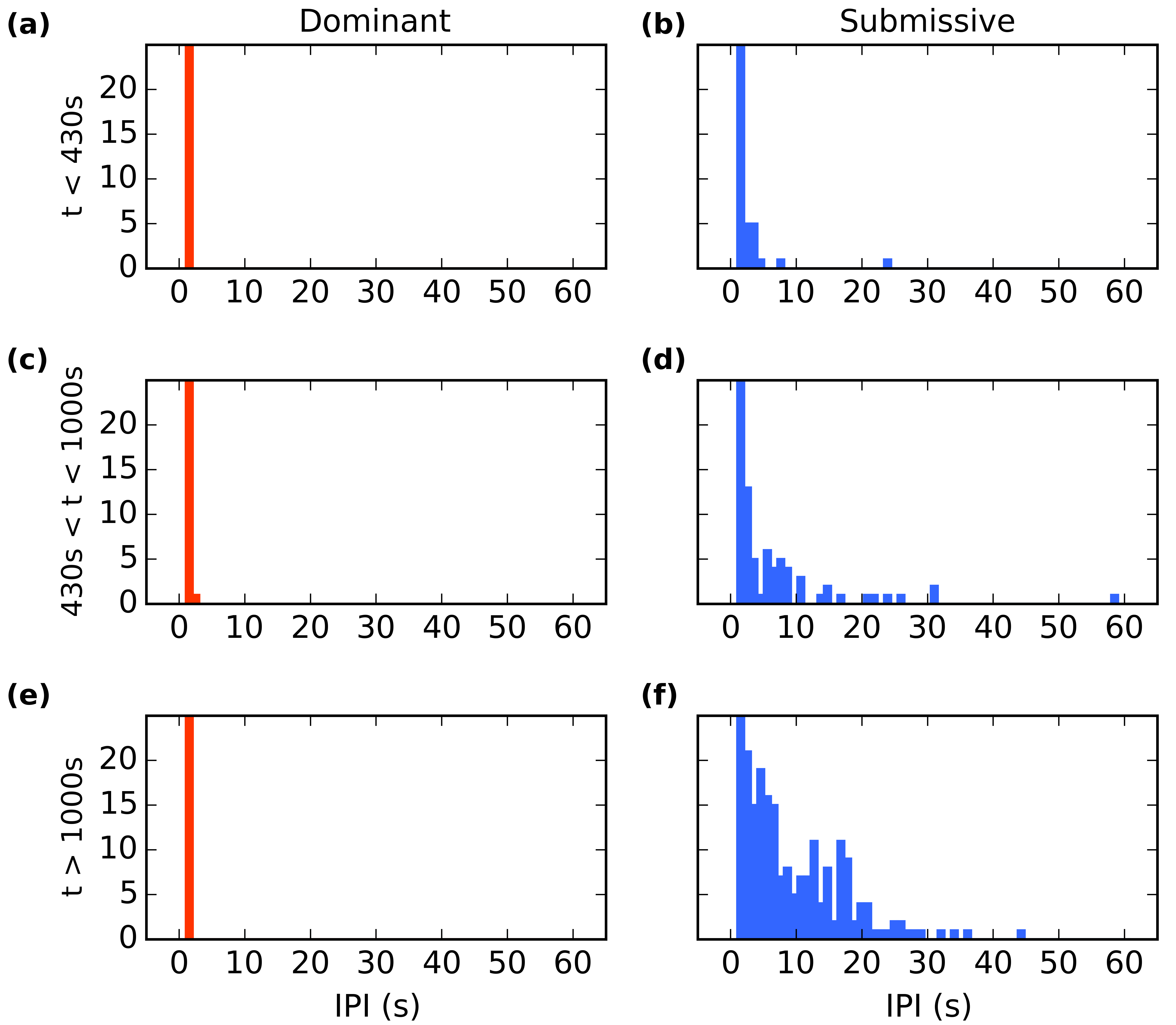}
\caption{Inter Pulse Intervals (IPI) on a coarser time scale. \textbf{Left:}
Dominant fish almost never shows chirps or moments of silence. \textbf{Right:}
Distribution of IPIs of the submissive fish becomes skewed as chirps and
moments of silence appear. \textbf{(a) and (b):} Histograms for the first 430
seconds. \textbf{(c) and (d):} Histograms for time between 430 and 1000
seconds. \textbf{(e) and (f):} Histograms for times greater than 1000 seconds.}
\label{fig:histCoarse}

\end{figure}


\begin{figure} \centering \includegraphics[width=1.2\textwidth]{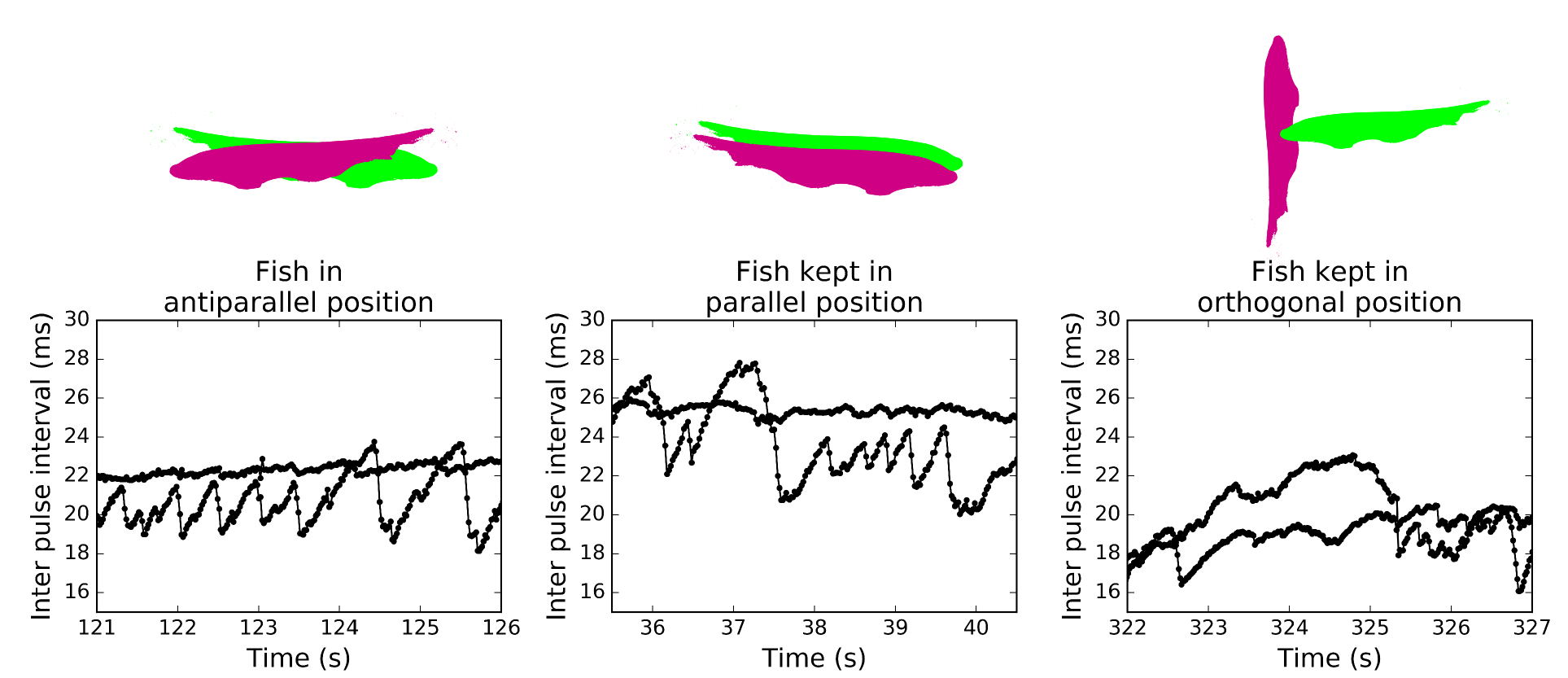}

  \caption{Fish manually held in several positions do not produce chirps when
in contact.  Representative sections of the IPI series are shown for each
relative position. Colors were not used because no direct dominance
relationship was observed in this experiment. \textbf{Top:} Relative positions
in which the fish were kept in physical contact. \textbf{Left:} Fish were
maintained in anti-parallel position. \textbf{Center:} Fish were maintained in
parallel position. \textbf{Right:} Fish were maintained in orthogonal
position.}


  \label{fig:esfrega}
\end{figure}



Chirps were automatically detected with $99.9$\% of accuracy
throughout all dyads tested. The accuracy in detecting chirps
saturated when the number of training examples $N$ was set to $10$\%
of the total number of chirps present in each time series, and using a
Random Forest composed of $200$ decision trees.  Remarkably, training
data from a single dyad allowed detection of all other dyads with the
same high accuracy, showing that our model learned general features of
regions with chirps without overfitting. In regions where chirps were
detected, updating the threshold for pulse detection improved the
accuracy of pulse discrimination (Figure
\ref{fig:chirp_importance}). Comparatively, ignoring chirps propagated
spurious detections in the pulse discrimination that propagated
throughout the time series. The current accuracy of the whole protocol
is higher than $97$\% (See Supplemental Material), which allows quick
manual corrections through the GUI.

Dominant and submissive fish changed their electrical behavior
differently over the course of the dominance contest (Figure
\ref{fig:series}). We discriminated pulses from six different fish
dyads during the first hour from their first encounter in the
aquarium, with precision of $0.5$~ms (Figure S2). The electrical behavior
of both fish go 
through a transient dynamics and stabilize after the first $16$min, as
seen in Figure \ref{fig:series} (see also Figure S3). Because most
dyads presented this transient behavior, we divided the IPIs time
series in three sections: early dominance contest, usually the first
$430$s; the dominance contest, between $430$s and $1000$s; and after
the dominance contest, which is after $1000$s. Although arbitrary, all
of our results are consistent with small changes in these
thresholds. During the early dominance contest, the distributions of
IPIs of both fish are very similar (Figure
\ref{fig:histFine}a,b). After a transient regime, when the submissive
fish starts changing its behavior, the IPIs of five of the submissive fish
changed significantly ($p-value \leq 0.01$ see table S2), while the average pulse rate
from the dominant fish remained unchanged on three of the six dyads (Figure
\ref{fig:histFine}c,d and $p-value \leq 0.01$ table S2). 
Dominant fish presented significantly less offs than submissive fish (Mann-Whitney U, 
$p<0.05$, see Figure S5), and no chirps were ever observed in dominant fish (Figure \ref{fig:resultsallpairs}).
Particularly, dyad $6$, the only one recorded in 2016 (June), has some periods where both fish presented a lower pulse rate.
Within these regions, no offs nor chirps were observed.
Although outside these regions this dyad
presents the same behavior in terms of offs and chirps as the other
dyads, it is the only pair that showed these low activity regions.  Due to these differences,
this dyad is not included on Figure
\ref{fig:resultsallpairs}. However, its data and statistical results
are found on Supplemental Material (Figures S3 and S6, and Tables S1 and S2).

Chirps and offs heavily skewed the distribution of IPIs of the submissive fish
(Figure \ref{fig:histCoarse}). 
The total duration of offs by submissive fish often represented a large portion of the whole time series 
(Figure \ref{fig:resultsallpairs}). 
There were cases when the submissive
fish emitted less than ten pulses between large periods of off (Figure S4 in the 
Supplemental Material). In some cases, the total amount of time of offs reached 
several minutes.

The IPI distribution when the fish is alone is always different from the
distribution during and after the conflict. This shows that the dynamic changes
observed in Figures \ref{fig:histFine} and \ref{fig:histCoarse} result from the
dispute for dominance, and are not due to intrinsic characteristics of the
fish.  

No chirps nor offs could be observed while fish were manually placed in
physical contact (Figure \ref{fig:esfrega}), which suggests that these
behaviors are not produced automatically by direct interference of the electric
organs.  Oscillatory behavior on the IPI timeseries was observed regardless of
their relative position (Figure \ref{fig:esfrega}). Interestingly, one of the
fish presented chirp in the interval between manipulations, suggesting that its
ability to generate chirps was not impaired by this manipulation.

\section{Discussion}

Electrocommunication in pulse-type electric fish (especially in many fish of
the genus \textit{Gymnotus} - \citealt{alvesGomes2013}) is achieved by a very
rich repertoire of electrical behaviors, most of which are present only when
fish are freely interacting. However, detecting and discriminating electrical
activity of interacting similar fish is a challenging task if their movements
are not severely restricted. We have developed a system to analyze pulse trains
of two electric fish freely swimming and interacting in the same aquarium. Our
system can automatically (i) detect chirps, and (ii) discriminate which fish
emitted each pulse. This technique allowed us to observe how the IPI
distributions of both fish changed over time in a fine-grained time scale.
While the submissive fish increased its average IPI when comparing the early
moments of the contest to its final, the dominant fish consistently kept its
IPI rate almost unchanged (with very similar average and variance - Figure
\ref{fig:resultsallpairs} The IPI time series are shown on Supplemental Material S3).


While the dominant fish maintained its pulse rate throughout the trials,
Submissive fish changed significantly their electrical behavior to different
average IPIs in almost all acquisitions (Figre \ref{fig:resultsallpairs}),
likely to avoid using the same average IPI of the dominant fish, as suggested
previously \citep{anaSilvaNonSex, anaSilva2015, westbyComparative}. Previous
studies also suggest that electric fish change its pulse rate in order to avoid
interference in electrolocation (\textit{Jamming Avoidance Response} -- JAR)
\citep{westbyJAR}. In addition, we observed signatures of JAR when the fish
were manually held during the attempt to produce chirps (Figure
\ref{fig:esfrega}).  However, when they are freely swimming, JAR are only
observed sparsely.
This could be evidence that when fish are not allowed to freely interact, they
express only a limited repertoire of EOD modulations, and JAR becomes more
evident.
Also, chirps were not observed by physical interference with the electric
organs, nor with artificial stimulation of a single fish. Although the
occurrence of chirps and offs may depend on other sensory and stereotypical
stimuli, such as bites and specific electrical patterns that naturally occur
during a dominance contest, this also suggests that these electrical behaviors
can also be used to communicate important messages that shape the dominance
contest and ultimately determine the dominance hierarchy.

Automating the detection of chirps proved extremely useful for pulse detection
and to minimize the laborious work of visual inspection through long IPI time
series. It improved the accuracy in pulse detection, avoiding the propagation
of errors in regions affected by chirps. Due to the generality of the electric
signature of chirps, it was possible to implement an efficient detector based
on a minimal dataset consisting of manually classified samples. Remarkably,
chirps in the time series of all dyads were automatically detected by using a
model trained over the data of a single dyad, which greatly improved the
applicability of this solution. With this methodology, long IPI time series
from interacting fish can be efficiently analyzed, which is fundamental for
investigating electrocommunication and complex hierarchical behaviors
\citep{mosqueiro2016non}.

Although the whole process is computationally intensive, we have optimized our
algorithms to allow them to run in a single off-the-shelf personal computer:
detection of chirps and pulses of an one hour experiment with two fish takes a
little more than one day, including manual corrections and inspection. Our code
have a friendly GUI and are freely available on GitHub \citep{gymnotools}. 
All our software is freely available \citep{gymnotools}. Data from one of our dyads is also available \citep{dados}.

\section{Funding}

Authors acknowledge useful discussions with Brenno Gustavo Barbosa. This work
was supported by the Brazilian agencies: Fundacão de Amparo à Pesquisa do
Estado de São Paulo – FAPESP (www.fapesp.br), Coordenação de Aperfeiçoamento de
Pessoal de Nível Superior – CAPES (www.capes.gov.br), and Conselho Nacional de
Desenvolvimento Científico e Tecnológico – CNPq (www.cnpq.br). TS Mosqueiro
acknowledges partial support from CNPq grant 234817/2014-3. The funders had no
role in study design, data collection and analysis, decision to publish, or
preparation of the manuscript.

\section{References}

\bibliographystyle{plainnat} \bibliography{references}

\end{document}